\documentclass[conference]{IEEEtran}
\IEEEoverridecommandlockouts

\usepackage[noadjust]{cite}

\usepackage[utf8]{inputenc}
\usepackage{booktabs}
\usepackage{multicol}
\usepackage{graphicx}
\usepackage{multirow}
\usepackage{amssymb,amsmath,bm}
\usepackage{enumitem}
\usepackage{soul,color}
\usepackage{array}
\usepackage{footnote}
\makesavenoteenv{table}
\usepackage{hyperref}
\usepackage{color}
\usepackage{graphicx}
\usepackage{subfigure}
\usepackage{xcolor}
\PassOptionsToPackage{hyphens}{url}\usepackage{hyperref}

\usepackage[left]{lineno}
\usepackage[utf8]{inputenc}
\usepackage{amsmath,bm}
\usepackage[ruled, linesnumbered]{algorithm2e}
\newcommand{\bs}[1]{\boldsymbol{-1}} 
\newcommand{\addcomment}[1]{\hl{\textbf{NOTE:}}\hl{\emph{-1}}}

\usepackage{float}

\def\model{NEST~}

\title{NEST: Self-supervised Fast Conformer as All-purpose Seasoning to Speech Processing Tasks}

\author{
\begin{tabular}{c} 
He Huang, Taejin Park, Kunal Dhawan, Ivan Medennikov, Krishna C. Puvvada, \\
Nithin Rao Koluguri, Weiqing Wang, Jagadeesh Balam, Boris Ginsburg \\
\textit{NVIDIA, Santa Clara, CA, USA}\\
\end{tabular} 
}

\begin{document}
%
\maketitle

\def\modelfullname{\emph{NeMo Encoder for Speech Tasks} (\textbf{NEST})}


\begin{abstract}
Self-supervised learning (SSL) has been proved to benefit a wide range of speech processing tasks, such as speech recognition/translation, speaker verification and diarization, etc. 
However, most of current speech SSL approaches are computationally expensive. 
In this paper, we introduce a simplified and more efficient SSL framework, termed as \modelfullname. Specifically, we adopt the FastConformer architecture with 8x sub-sampling rate, which is faster than Transformer or Conformer architectures. Instead of clustering-based quantization, we use fixed random projection for its simplicity and effectiveness. We also implement a generalized noisy speech augmentation that teaches the model to disentangle the main speaker from noise or other speakers. Experiments show that \model improves over existing self-supervised models and achieves new state-of-the-art performance on a variety of speech processing tasks, such as speech recognition/translation, speaker diarization, spoken language understanding, etc. Code and checkpoints are publicly available via NVIDIA NeMo framework\footnote{\texttt{https://github.com/NVIDIA/NeMo}}\footnote{\texttt{https://huggingface.co/nvidia/ssl\_en\_nest\_large\_v1.0}}\footnote{\texttt{https://huggingface.co/nvidia/ssl\_en\_nest\_xlarge\_v1.0}}.

\end{abstract}
%
\begin{IEEEkeywords}
self-supervised learning, speech recognition, speaker diarization, spoken language understanding
\end{IEEEkeywords}

\section{Introduction}


Most recent speech self-supervised models are inspired by the BERT~\cite{devlin2018bert} model, which learn text token embedding by predicting the target of the masked positions given the context of the unmasked ones. 
Among them are two main streams of \emph{contrastive} and \emph{predictive} models. The \emph{contrastive} approach~\cite{baevski2020wav2vec,sadhu2021wav2vec_c,baevski2019vq_wav2vec,jiang2020speech_simclr} quantizes the speech features into a set of target feature \emph{vectors} and trains with a contrastive loss using the positive and negative target features. Meanwhile, the \emph{predictive} approach~\cite{hsu2021hubert,chiu2022best_rq,chen2022wavlm,chen2024xeus} quantizes the speech features into \emph{tokens} and train with masked token prediction loss as in BERT~\cite{devlin2018bert}. In addition to the two approaches, some works also learn from the masked auto-encoding~\cite{he2022masked} approach and train speech self-supervised models with a \emph{reconstruction} objective~\cite{baevski2022data2vec,jiang2020speech_simclr}.

One representative work of \emph{contrastive} models is Wav2vec-2.0~\cite{baevski2020wav2vec}, which demonstrates initializing ASR models from SSL checkpoints can outperform previous semi-supervised and train-from-scratch ASR models. 
Later, Wav2vec-C~\cite{sadhu2021wav2vec_c} improves over Wav2vec-2.0 by adding a consistency loss to reconstruct the quantized embedding, similar to VQ-VAE~\cite{van2017vq_vae}. 
XLS-R~\cite{babu2021xls} extends Wav2vec-2.0 to multilingual setting and shows impressive performance on multilingual speech recognition and translation.

\begin{figure}[!t]
    \centering
    \includegraphics[width=0.9\linewidth]{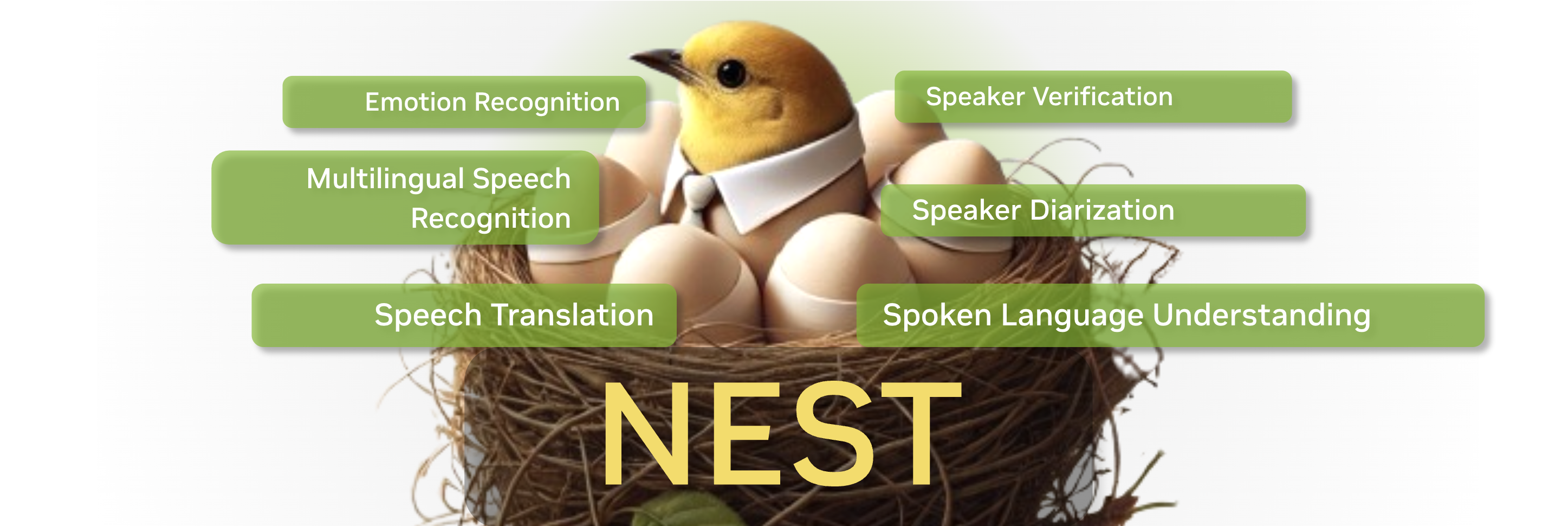}
    \vspace{-2px}
    \caption{NEST serves as a bird nest that incubates the variety of speech task models.}
    \label{fig:nest_tasksl}
    \vspace{-4ex}
\end{figure}

HuBERT~\cite{hsu2021hubert}, as a pioneer work of the \emph{predictive} approach, generates the target tokens by running k-means clustering on the middle layer features extracted from another SSL model that is pretrained for a small number of steps. Then, W2v-BERT~\cite{chung2021w2v_bert} proposes to combine the training objectives of both Wav2vec-2.0~\cite{baevski2020wav2vec} and HuBERT~\cite{hsu2021hubert} by applying contrastive loss on the middle layer output and predictive loss on the final output layer. Later, BEST-RQ~\cite{chiu2022best_rq} shows that the clustering based token generation can be replaced by simple fixed random-projection quantization, and this simple modification is able to match or outperform HuBERT on ASR.

\begin{figure*}[!tbh]
    \centering
    \includegraphics[width=0.9\linewidth]{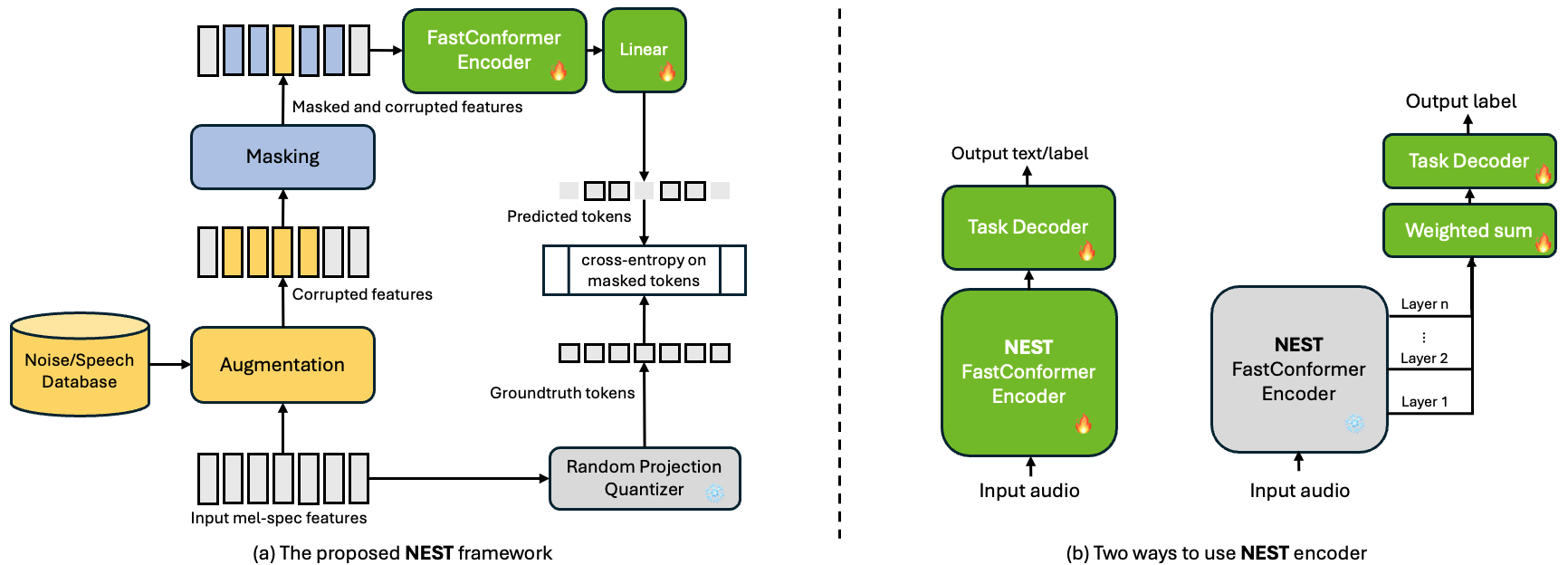}
    \caption{(a) The proposed \model framework for speech self-supervised learning. (b) Two ways to use \model encoder: (left) use as weight initialization for tasks that require more parameters (\textit{e.g.}, speech recognition); (right) learn weighted summation of features from different layers of the frozen \model for tasks that require less trainable parameters (\textit{e.g.,} speaker verification).}
    \label{fig:model}
\end{figure*}

In order to improve performance on speaker tasks, WavLM~\cite{chen2022wavlm} proposes a noisy speech augmentation technique and a \emph{denoising masked token prediction} objective, by adding a speech segment of a different speaker to the current speech and training the model to predict the target tokens generated using original clean speech. XEUS~\cite{chen2024xeus} further extends WavLM~\cite{chen2022wavlm} by adding a de-reverberation task and trains on multilingual data of 1M hours.

However, previous SSL models have notable limitations. First, several models~\cite{baevski2020wav2vec, hsu2021hubert, chen2022wavlm} employ a CNN-Transformer architecture with a relatively short frame length of 20 ms, which negatively impacts inference speed. Second, HuBERT-style quantization is highly computationally intensive, consuming up to 20\% of the total training time, as reported by XEUS~\cite{chen2024xeus}. Third, although BEST-RQ~\cite{chiu2022best_rq} uses Conformer~\cite{gulati2020conformer} encoder with 40ms frame length and simple random quantization, it lacks the ability to explicitly tell one speaker from another, which limits its performance on speaker tasks such as speaker diarization.


In this paper, we tackle all these challenges and bring the best practices from previous works, which constitute the proposed \modelfullname. Our contributions are summarized as follows:
\begin{itemize}
    \item A new speech self-supervised learning framework with a simplified and more streamlined design.
    \item Experiment results demonstrate that \model can help achieve SOTA performance on a variety of downstream tasks (ASR, AST, SD, SLU etc).
    \item Unlike previous SSL approaches that primarily focus on downstream tasks with very limited data, we also show that \model can benefit speech recognition and translation even when data is relatively larger.
    \item To the best of our knowledge, we are the first to show that SSL model trained on English data can also help improve speech recognition on other languages.
\end{itemize}

\section{Approach}


\subsection{Speech Encoder}
Current SOTA speech SSL models~\cite{chen2022wavlm,hsu2021hubert} mostly use transformer encoder~\cite{vaswani2017attention} or Conformer~\cite{gulati2020conformer} as speech encoder, which have either 20ms or 40ms frame length. Here we choose the more efficient FastConformer~\cite{rekesh2023fast} which applies 8x convolutional sub-sampling on the input Mel-spectrogram before the following FastConformer layers, resulting in an 80ms frame length that can significantly reduce the sequence length to be processed by self-attention layers. 

\subsection{Speech Augmentation}
We augment the input speech with random noise or speech of another speaker, similar to the techniques proposed in WavLM~\cite{chen2022wavlm}. However, we generalize the augmentation in three ways: (1) the length of augmentation audio is sampled between 0.4 and 0.6 of the primary audio length, instead of a fixed 0.5 ratio~; (2) the length of augmentation audio is randomly split into 1, 2 or 3 segments with uniform probability, such that the augmentation is scattered to different positions of the primary audio; (3) instead of using single negative speaker, for each segment with speaker augmentation, we randomly select a different speaker from other speakers in the same batch, such that there can be more speakers in the resulted audios.

\begin{table*}[!tbh]
\centering
\caption{Results on SUPERB~\cite{yang2021superb} benchmark for multi-task evaluation on SSL speech encoders.}
\vspace{-2ex}
\label{tab:superb}
\resizebox{\textwidth}{!}{
\begin{tabular}{l|r|c|ccc|ccc|c}
\toprule
\multicolumn{1}{c|}{\multirow{2}{*}{Model}} & \multicolumn{1}{c|}{\multirow{2}{*}{Params}} & \multicolumn{1}{c|}{\multirow{2}{*}{SSL Data (hrs)}} & \multicolumn{3}{c|}{Speaker} & \multicolumn{3}{c|}{Content} & ParaLinguistics \\
\multicolumn{1}{c|}{} & \multicolumn{1}{c|}{} & \multicolumn{1}{c|}{} & \multicolumn{1}{c|}{SID (Acc $\uparrow$)} & \multicolumn{1}{c|}{SV (EER $\downarrow$)} & SD (DER $\downarrow$) & \multicolumn{1}{c|}{PR (PER $\downarrow$)} & \multicolumn{1}{c|}{ASR (WER $\downarrow$)} & KS (Acc $\uparrow$) & ER (Acc $\uparrow$) \\ \hline \hline
WavLM-base++~\cite{chen2022wavlm} & 95M & En-96K & \multicolumn{1}{c|}{86.84} & \multicolumn{1}{c|}{4.26} & 4.07 & \multicolumn{1}{c|}{4.07} & \multicolumn{1}{c|}{5.59} & 96.69 & 67.98 \\  
WavLM-large~\cite{chen2022wavlm} & 316M & En-96K & \multicolumn{1}{c|}{95.25} & \multicolumn{1}{c|}{4.04} & 3.47 & \multicolumn{1}{c|}{3.09} & \multicolumn{1}{c|}{3.44} & 97.40 & 70.03 \\  
XEUS~\cite{chen2024xeus} & 577M & MulLing-1M & \multicolumn{1}{c|}{91.70} & \multicolumn{1}{c|}{4.16} & 3.11 & \multicolumn{1}{c|}{3.21} & \multicolumn{1}{c|}{3.34} & \textbf{98.32} & \textbf{71.08} \\ \hline \hline
NEST-L & 115M & En-100K & \multicolumn{1}{c|}{94.94} & \multicolumn{1}{c|}{3.85} & 2.28 & \multicolumn{1}{c|}{1.95} & \multicolumn{1}{c|}{3.49} & 96.85 & 68.12 \\ 
NEST-XL & 600M & En-100K & \multicolumn{1}{c|}{\textbf{95.76}} & \multicolumn{1}{c|}{\textbf{2.49}} & \textbf{1.89} & \multicolumn{1}{c|}{\textbf{1.80}} & \multicolumn{1}{c|}{\textbf{3.19}} & 97.11 & 69.94 \\ \bottomrule
\end{tabular}}
\vspace{-3ex}
\end{table*}

\begin{table*}[]
\centering
\caption{Results on multi-lingual ASR with punctuation and capitalization. Performance is evaluated by word error rate ({WER}) including native punctuation and capitalization from the source datasets. \underline{Underline} indicates the second best performance.}
\vspace{-2ex}
\label{tab:asr}
\resizebox{\textwidth}{!}{
\begin{tabular}{l|r|c|cc|cc|cc|cc|c}
\toprule
\multicolumn{1}{c|}{\multirow{2}{*}{Model}} & \multirow{2}{*}{Params} & \multirow{2}{*}{Data (hrs)} & \multicolumn{2}{c|}{En} & \multicolumn{2}{c|}{De} & \multicolumn{2}{c|}{Es} & \multicolumn{2}{c|}{Fr} & \multirow{2}{*}{Avg} \\
\multicolumn{1}{c|}{} &  &  & \multicolumn{1}{c|}{MCV16.1} & Voxpopuli & \multicolumn{1}{c|}{MCV16.1} & Voxpopuli & \multicolumn{1}{c|}{MCV16.1} & Voxpopuli & \multicolumn{1}{c|}{MCV16.1} & Voxpopuli &  \\ \hline \hline
SeamlessM4T-medium-v1~\cite{barrault2023seamless} & 1.2B & 4M & \multicolumn{1}{c|}{14.20} & 10.02 & \multicolumn{1}{c|}{11.25} & 16.20 & \multicolumn{1}{c|}{11.43} & 12.01 & \multicolumn{1}{c|}{17.34} & 12.49 & 13.11 \\ 
SeamlessM4T-large-v2~\cite{barrault2023seamless} & 2.3B & 4M & \multicolumn{1}{c|}{\textbf{11.13}} & 7.77 & \multicolumn{1}{c|}{\textbf{7.53}} & 13.39 & \multicolumn{1}{c|}{\underline{8.67}} & 10.53 & \multicolumn{1}{c|}{\textbf{14.37}} & 10.13 & \textbf{10.44} \\ 
Whisper-large-v3~\cite{radford2023whisper} & 1.5B & 5M & \multicolumn{1}{c|}{15.73} & 13.42 & \multicolumn{1}{c|}{9.24} & 21.41 & \multicolumn{1}{c|}{10.95} & 14.31 & \multicolumn{1}{c|}{17.35} & 13.58 & \underline{14.49} \\ 
Canary-1b~\cite{puvvada2024less} & 1B & 86k & \multicolumn{1}{c|}{\underline{12.46}} & \textbf{7.52} & \multicolumn{1}{c|}{8.71} & 15.32 & \multicolumn{1}{c|}{\textbf{8.28}} & \underline{9.56} & \multicolumn{1}{c|}{\underline{15.46}} & \textbf{8.78} & 10.76 \\ 
FastConformer-XL-hybrid (ASR init) & 600M & 14K & \multicolumn{1}{c|}{16.78} & 8.21 & \multicolumn{1}{c|}{9.17} & \underline{12.69} & \multicolumn{1}{c|}{9.75} & 10.19 & \multicolumn{1}{c|}{17.42} & 9.89 & 11.76 \\ \hline \hline
NEST-XL-hybrid & 600M & 14K & \multicolumn{1}{c|}{14.43} & \underline{7.58} & \multicolumn{1}{c|}{\underline{8.07}} & \textbf{11.83} & \multicolumn{1}{c|}{8.70} & \textbf{9.27} & \multicolumn{1}{c|}{16.18} & \underline{9.74} & {10.72} \\ \bottomrule
\end{tabular}
}\vspace{-4ex}
\end{table*}

\subsection{Speech Quantization}

We use BEST-RQ~\cite{chiu2022best_rq} for speech quantization. Specifically, we employ a single randomly initialized and frozen codebook of 8192 vocabulary and 16 dimension features. A randomly initialized and frozen linear layer is applied to the input Mel-spectrogram features to project them into the same dimension as the codebook, then a nearest neighbor search is applied to obtain the target tokens. Since there is an 8x subsampling, we channel-concatenate the features for each 8 consecutive frames before feeding into the linear layer, such that the lengths for the target tokens and input features are equal.

\subsection{Feature Masking}
We employ a random block-wise masking mechanism on the input Mel-spectrogram features, where each frame in the input has a probability $p_m$ as being selected as the start of a masking block. After randomly  selecting a set of starting frames, we mask $l_m$ consecutive frames for each of the starting frames. Note that there could be overlapping between two masked blocks, which allows for arbitrary lengths in the resulting masked segments that do not overlap with each other. We use $p_m=0.01$ and $l_m=40$ in all our experiments.

\subsection{Training}

Since masking is performed before the convolutional sub-sampling, there is a mismatch in the lengths between the predictions and masks. To match the sequence lengths, masks are averaged for every 8 frames, then apply threshold of 0.9 to select frames to be taken into loss calculation. Cross-entropy loss is applied on selected positions determined by the averaged input masks.



\section{Experiments}

\subsection{Dataset and Settings}
We train the NEST-L (115M) and NEST-XL (600M) models using 100K hours of English speech data, including 60K hours from LibriLight~\cite{kahn2020librilight}, 24K hours from English subset of Voxpopuli~\cite{wang2021voxpopuli}, and about 20K hours sampled data from the combination of Fisher~\cite{cieri2004fisher}, Switchboard~\cite{switchboard}, WSJ~\cite{wsj}, NSC~\cite{koh2019nsc}, People's Speech~\cite{galvez2021people}. The audios for speech augmentation are randomly selected within each batch for each sample, and we use non-vocal noise audios from MUSAN~\cite{snyder2015musan} and Freesound~\cite{fonseca2017freesound}. We train the models with global batch size of 2048 samples for about 800K steps on 128 NVIDIA A100 GPUs, with Noam annealing~\cite{vaswani2017attention}, peak learning rate of 0.004, weight decay of 1e-3, gradient clipping 1.0 and warm-up of 25K steps. We set the speech augmentation probability as 0.2, among which we set noise and speech augmentation probabilities as 0.1 and 0.9.


\subsection{Results on SUPERB Multi-task Speech Processing}

We evaluate our model's performance on the SUPERB~\cite{yang2021superb} benchmark for multi-task evaluation on self-supervised speech models. For speech recognition (ASR), phoneme recognition (PR) and speaker diarization (SD) tasks, we use the architecture in the left part of Figure~\ref{fig:model}(b) and a simple linear layer as the task decoder. We train ASR and PR with CTC~\cite{graves2006connectionist} loss, while the SD task is trained with permutation invariant loss (PIL)~\cite{fujita2019end}. For speaker identification/verification (SID/SV), keyword spotting (KS) and emotion recognition (ER) tasks, we resort to the architecture presented in the right part of Figure~\ref{fig:model}(b), and use the ECAPA-TDNN-small~\cite{desplanques2020ecapa} as the task decoder. We following the same train/val/test splits as in the SUPERB~\cite{yang2021superb} and train the models for 100 epochs.

As presented in Table~\ref{tab:superb}, our NEST-L model is able to outperform WavLM-base++~\cite{chen2022wavlm} with similar size of parameters on all tasks, and also outperforms WavLM-large~\cite{chen2022wavlm} that is 3x as large on speaker verification (SV), speaker diarization (SD) and phoneme recognition (PR). When compared with the XEUS~\cite{chen2024xeus} model that is trained on 10x data, we can see that our NEST-XL model is still able to achieve better performance on all speaker and content tasks, with especially large improvements on speaker verification, speaker diarization and phoneme recognition. Overall, we are able to achieve new state-of-the-art results on SID, SV, SD, PR and ASR tasks compared with WavLM~\cite{chen2022wavlm} that has similar data size as well as XEUS~\cite{chen2024xeus} that is trained on much large data, demonstrating the effectiveness of \model when applied on various downstream speech processing tasks.

\begin{table*}[!t]
\caption{Results on speech translation from English to German, French and Spanish. BLEU score is used as the metric, while punctuation and capitalization are included in metric calculation. \underline{Underline} indicates second best performance.}
\vspace{-2ex}
\label{tab:ast}
\resizebox{\textwidth}{!}{
\begin{tabular}{l|r|c|ccc|ccc|ccc|ccc}
\toprule
\multicolumn{1}{c|}{\multirow{2}{*}{Model}} & \multirow{2}{*}{Params} & \multirow{2}{*}{Data (hrs)} & \multicolumn{3}{c|}{Europarl} & \multicolumn{3}{c|}{mExpresso} & \multicolumn{3}{c|}{Fleurs} & \multicolumn{3}{c}{Average} \\
\multicolumn{1}{c|}{} &  &  & \multicolumn{1}{c|}{En$\rightarrow$De} & \multicolumn{1}{c|}{En$\rightarrow$Es} & En$\rightarrow$Fr & \multicolumn{1}{c|}{En$\rightarrow$De} & \multicolumn{1}{c|}{En$\rightarrow$Es} & En$\rightarrow$Fr & \multicolumn{1}{c|}{En$\rightarrow$De} & \multicolumn{1}{c|}{En$\rightarrow$Es} & En$\rightarrow$Fr & \multicolumn{1}{c|}{En$\rightarrow$De} & \multicolumn{1}{c|}{En$\rightarrow$Es} & En$\rightarrow$Fr \\ \hline \hline
SeamlessM4T-medium & 1.2B & 4M & \multicolumn{1}{c|}{28.03} & \multicolumn{1}{c|}{38.44} & \underline{30.50} & \multicolumn{1}{c|}{9.65} & \multicolumn{1}{c|}{16.23} & 8.64 & \multicolumn{1}{c|}{28.30} & \multicolumn{1}{c|}{21.05} & 37.36 & \multicolumn{1}{c|}{21.99} & \multicolumn{1}{c|}{25.24} & 25.50 \\ 
SeamlessM4T-v2-large & 2.3B & 4M & \multicolumn{1}{c|}{19.96} & \multicolumn{1}{c|}{32.32} & 23.33 & \multicolumn{1}{c|}{21.48} & \multicolumn{1}{c|}{34.89} & 26.04 & \multicolumn{1}{c|}{\textbf{33.17}} & \multicolumn{1}{c|}{\textbf{23.72}} & \textbf{43.05} & \multicolumn{1}{c|}{24.87} & \multicolumn{1}{c|}{30.31} & 30.80 \\ 
Canary-1B~\cite{puvvada2024less} & 1B & 86K & \multicolumn{1}{c|}{\textbf{32.53}} & \multicolumn{1}{c|}{\textbf{40.84}} & \textbf{30.65} & \multicolumn{1}{c|}{\textbf{23.83}} & \multicolumn{1}{c|}{\textbf{35.73}} & \textbf{28.28} & \multicolumn{1}{c|}{\underline{32.15}} & \multicolumn{1}{c|}{\underline{22.66}} & \underline{40.76} & \multicolumn{1}{c|}{\textbf{29.50}} & \multicolumn{1}{c|}{\textbf{33.07}} & \textbf{33.23} \\ \hline \hline 
NEST-XL-Transformer & 1B & 42K & \multicolumn{1}{c|}{\underline{30.87}} & \multicolumn{1}{c|}{\underline{39.95}} & 30.01 & \multicolumn{1}{c|}{\underline{22.82}} & \multicolumn{1}{c|}{\underline{34.92}} & \underline{27.99} & \multicolumn{1}{c|}{29.50} & \multicolumn{1}{c|}{22.61} & 39.27 & \multicolumn{1}{c|}{\underline{27.73}} & \multicolumn{1}{c|}{\underline{32.51}} & \underline{32.42} \\ \bottomrule
\end{tabular}
}
\vspace{-4ex}
\end{table*}

 
\begin{table}[]
\centering
\caption{DER results on speaker diarization. \underline{Underline} indicates the second best performance. Starred(*) systems are not end-to-end systems which involve clustering steps.}
\label{tab:sd}
\resizebox{\columnwidth}{!}{
\begin{tabular}{l|c|c|c|c}
\toprule
\multicolumn{1}{c|}{\multirow{2}{*}{Model}} & DIHARD3-eval & \multicolumn{3}{c}{CALLHOME-part2} \\
\multicolumn{1}{c|}{} & \begin{tabular}[c]{@{}c@{}} $\leq$4 speakers, \\ collar=0.0\end{tabular} & \multicolumn{1}{c|}{\begin{tabular}[c]{@{}c@{}}2 speakers, \\ collar=0.25\end{tabular}} & \multicolumn{1}{c|}{\begin{tabular}[c]{@{}c@{}}3 speakers, \\ collar=0.25\end{tabular}} & \begin{tabular}[c]{@{}c@{}}4 speakers, \\ collar=0.25\end{tabular} \\ \hline \hline
EEND-EDA~\cite{horiguchi2022eend_eda,horiguchi2022eend_gla} & 15.55 & \multicolumn{1}{c|}{7.83} & \multicolumn{1}{c|}{12.29} & 17.59 \\ 
*WavLM-L+EEND-VC~\cite{chen2022wavlm}  & N/A & 6.46  & \underline{10.69} & \textbf{11.84} \\
*EEND-GLA-small~\cite{horiguchi2022eend_gla} &\textbf{14.39} & 6.94 & 11.42 & \underline{14.49}  \\
*NeMo MSDD~\cite{park2022msdd} & 29.40 & \multicolumn{1}{c|}{11.41} & \multicolumn{1}{c|}{16.45} & 19.49 \\ \hline \hline
RandFC-L-MLP & 21.71 & \multicolumn{1}{c|}{11.60} & \multicolumn{1}{c|}{15.89} & 21.38 \\ 
NEST-L-MLP & 16.83 & \multicolumn{1}{c|}{7.88} & \multicolumn{1}{c|}{11.71} & 20.22 \\ \hline \hline
RandFC-L-Sortformer-HL~\cite{park2024sortformer} & 18.93 & \multicolumn{1}{c|}{9.39} & \multicolumn{1}{c|}{13.56} & 20.15 \\ 
NEST-L-Sortformer-HL~\cite{park2024sortformer} & 16.28 & \multicolumn{1}{c|}{\underline{6.49}} & \multicolumn{1}{c|}{10.01} & 14.14 \\ 
NEST-L-Sortformer-HL-PP~\cite{park2024sortformer} & \underline{14.76} & \multicolumn{1}{c|}{\textbf{5.87}} & \multicolumn{1}{c|}{\textbf{8.46}} & 12.59 \\ 
\bottomrule
\end{tabular}
}
\vspace{-4ex}
\end{table}

\subsection{Results on Multi-lingual Speech Recognition}

Besides multi-task evaluation, we also study if an SSL model trained on single language can help other languages. To this end, we train an ASR model on four different languages: English (En), German (De), French (Fr), Spanish (Es). Specifically, we train an ASR model using NEST-XL as weight initialization and the hybrid-CTC-RNNT loss~\cite{noroozi2024stateful}. The training data comprises of 8.5K hours of English speech (MCV~\cite{commonvoice:2020}, MLS~\cite{pratap2020mls}, Voxpopuli~\cite{wang2021voxpopuli}, SPGI~\cite{o2021spgispeech}, Europarl~\cite{iranzo2020europarl}, LibriSpeech~\cite{panayotov2015librispeech}, NSC1~\cite{koh2019nsc}, Fisher~\cite{cieri2004fisher}), 2.5K hours of German speech (MCV, MLS, Voxpopuli), 1.4K hours of Spanish speech (MCV, MLS, Voxpopuli) and 1.9K hours of French speech (MCV, MLS, Voxpopuli). For baselines, we train another model using an English ASR model~\cite{stt_en_fastconformer_transducer_xlarge} as weight initialization, and also include some of the best ASR models like Whisper~\cite{radford2023whisper}, SeamlessM4T~\cite{barrault2023seamless} and Canary~\cite{puvvada2024less}. We run all models with the same beam size 5 with no language models on test sets of MCV-16.1~\cite{commonvoice:2020} and Voxpopuli~\cite{wang2021voxpopuli}.

From the last two rows of Table~\ref{tab:asr}, we can see that \model can help achieve better WER on all datasets than the model with ASR pretrained initialization, which shows that \model can help improve ASR performance on languages that is not seen during SSL pretraining. In addition, when compared with other SOTA ASR models (Whisper~\cite{radford2023whisper}, SeamlessM4T~\cite{barrault2023seamless}, Canary~\cite{puvvada2024less}) trained with much more parameters and data, we are still able to match the performance of Canary~\cite{puvvada2024less} on averaged WER across all languages. On some of the datasets, although there is still a gap between our model's performance and that of the SOTA models trained with much more data, we can still see that \model can be used as an efficient way to obtain good ASR performance comparable to models trained on massive datasets.

\subsection{Results on Speech Translation}

We further study how \model can help speech-to-text translation (AST) and present the results in Table~\ref{tab:ast}. We use the same model architecture and training procedure as proposed in Canary~\cite{puvvada2024less}, while the training data contains 42K hours of English ASR data with machine generated translation~\cite{MegatronNMTAnyEn2024} from English (En) to German (De), French (Fr) and Spanish (Es) text. We compare our model with other SOTA AST models (\emph{e.g.}, SeamlessM4T~\cite{barrault2023seamless} and Canary~\cite{puvvada2024less}) on Europarl~\cite{iranzo2020europarl}, mExpresso~\cite{mexpresso} and FLEURS~\cite{conneau2023fleurs} test sets. Given the same number of parameters, due to much less training data, there is still a gap between Canary~\cite{puvvada2024less} and our model on all evaluated datasets. Also, given that Canary~\cite{puvvada2024less} is initialized with a multi-lingual ASR encoder that is pretrained on all of the evaluated languages, it is expected that Canary performs better than the English-only  \model initialization. Nonetheless, our model is able to outperform SeamlessM4T~\cite{barrault2023seamless} and achieves the second best average BLEU scores on En$\rightarrow$De, En$\rightarrow$Es and En$\rightarrow$Fr translations, showing that \model is able to help achieve impressive AST performance with less data.

\subsection{Results on Speaker Diarization}

To assess the impact of \model on speaker diarization, we train two variants of end-to-end diarization models:  (1) simple two-layer multi-layer-perceptron (MLP) on top of FastConformer encoder and train with PIL; (2) more sophisticated Sortformer~\cite{park2024sortformer} hybrid loss (HL) model with post processing (PP)\footnote{Post-processing parameters were tuned separately for DIHARD3 and CALLHOME on corresponding training parts.} with 18 layers of transformer on top of the encoder. We also apply {NEST} and random initialization to both models for comparison. For training data, we use a combination of 2030 hours of real data (Fisher English~\cite{cieri2004fisher}, AMI Mix-Headset train+dev~\cite{ami}, ICSI~\cite{icsi}, DIHARD3 dev~\cite{ryant2020third},  VoxConverse v0.3~\cite{voxconv}, AISHELL-4~\cite{fu2021aishell}, CALLHOME-part1\footnote{We follow splits from the Kaldi x-vector recipe~\cite{callhome2} by using \texttt{part1} for training and  \texttt{part2} for evaluation.}~\cite{callhome}) and 5150 hours of simulated data (composed from LibriSpeech~\cite{panayotov2015librispeech} and SRE~\cite{doddington2000nist,nist}) generated by the NeMo speech data simulator~\cite{park2023property}. We evaluate models' performance on DIHARD3-eval~\cite{ryant2020third} and CALLHOME-part2~\cite{callhome}.

As shown in Table~\ref{tab:sd}, by comparing RandFC-L-MLP with NEST-L-MLP, and RandFC-L-Sortformer-HL with NEST-L-Sortformer-HL, we can see that \model provides significant improvements ($1\sim5\%$ absolute DER in different settings) over randomly initialized encoder, which demonstrates the effectiveness of NEST in speaker diarization task.
We can also see that Sortformer~\cite{park2024sortformer} with \model initialization is able to achieve second best results on DIHARD3 eval set when using postprocessing (PP), and it also achieves new SOTA results on 2 and 3 speaker settings of CALLHOME-part2 within all compared methods. Among end-to-end methods, NEST-L-Sortformer-HL-PP is able to outperform EEND-EDA~\cite{horiguchi2022eend_eda} on all test sets, while RandFC-L-Sortformer-HL lags behind, showing that \model is essential for achieving SOTA results in end-to-end speaker diarization.

\begin{table}[]
\centering
\caption{Results on SLURP~\cite{bastianelli2020slurp} benchmark for end-to-end speech joint intent detection and slot filling.}
\label{tab:slurp}
\resizebox{\columnwidth}{!}{
\begin{tabular}{l|c|c|c|c|c}
\toprule
\multicolumn{1}{c|}{Model} & \multicolumn{1}{c|}{\begin{tabular}[c]{@{}c@{}}SSL \\ Data (hrs)\end{tabular}} & \begin{tabular}[c]{@{}c@{}}Intent \\ Acc\end{tabular} & \begin{tabular}[c]{@{}c@{}}SLURP-\\ Precision\end{tabular} & \begin{tabular}[c]{@{}c@{}}SLURP-\\ Recall\end{tabular} & \begin{tabular}[c]{@{}c@{}}SLURP-\\ F1\end{tabular} \\ \hline \hline
SpeechBrain-Hubert-large~\cite{wang2021fine} & LL-60K & { 89.37} & { 80.54} & { 77.44} & { 78.96} \\ 
ESPnet-Conformer~\cite{arora2022espnet} & N/A & { 86.30} & { N/A} & { N/A} & { 71.40} \\ 
Open-BEST-RQ~\cite{whetten2024open} & LS-960 & { 74.80} & { N/A} & { N/A} & { N/A} \\ 
Wav2vec-CTI-RoBERTa~\cite{seo2022integration} & LS-960 & { 86.92} & { N/A} & { N/A} & { 74.66} \\ 
NeMo-SSL-FC-Trans-L~\cite{huang2023leveraging} & LL-60K & { 89.40} & { 77.90} & { 76.65} & { 77.22} \\ \hline \hline
NEST-L-Transformer & En-100K & {\textbf{89.79}} & { 80.55} & { \textbf{78.70}} & { 79.61} \\ 
NEST-XL-Transformer & En-100K & { 89.04} & { \textbf{82.35}} & { 78.36} & { \textbf{80.31}} \\ \bottomrule
\end{tabular}
}
\vspace{-4ex}
\end{table}

\subsection{Results on Spoken Language Understanding}

For spoken language understanding, we focus on the \emph{joint intent detection and slot filling} task and evaluate our model's performance using the SLURP~\cite{bastianelli2020slurp} dataset. Specifically, we attach a transformer decoder to the \model encoder, and use the same hyper-parameter setting as in  NeMo-SLU~\cite{huang2023leveraging}. We compare with other SSL-based end-to-end SLU models and show the results in Table~\ref{tab:slurp}. For fair comparison, we do not include the ASR pretrained baseline~\cite{huang2023leveraging} as we focus on SSL.

As we can see, among all SSL-based SLU models, using \model as speech encoder can help achieve the best performance on both intent detection accuracy and slot filling F1 scores. We also notice that scaling up from NEST-L to NEST-XL does bring some improvement on precision score on slot filling, but do not have significant effects on other metrics. In addition, compared with the NeMo-SSL-FC-Trans-L~\cite{huang2023leveraging} baseline, we can see a more than 2\% absolute improvement on F1 score by merely replacing the SSL speech encoder with \model while keeping other hyper-parameters the same, which demonstrates the instant benefits that \model can bring to existing speech processing models.

\section{Conclusion}

In this paper, we introduced a simplified and efficient speech self-supervised learning framework termed NEST, and extensive experiments on multiple speech processing tasks show that the \model framework can help achieve state-of-the-art performance. Code, configurations and checkpoints are available through NVIDIA NeMo framework.

\bibliographystyle{IEEEtran}
\bibliography{reference}

\begin{thebibliography}{10}
\providecommand{\url}[1]{#1}
\csname url@samestyle\endcsname
\providecommand{\newblock}{\relax}
\providecommand{\bibinfo}[2]{#2}
\providecommand{\BIBentrySTDinterwordspacing}{\spaceskip=0pt\relax}
\providecommand{\BIBentryALTinterwordstretchfactor}{4}
\providecommand{\BIBentryALTinterwordspacing}{\spaceskip=\fontdimen2\font plus
\BIBentryALTinterwordstretchfactor\fontdimen3\font minus \fontdimen4\font\relax}
\providecommand{\BIBforeignlanguage}[2]{{%
\expandafter\ifx\csname l@#1\endcsname\relax
\typeout{** WARNING: IEEEtran.bst: No hyphenation pattern has been}%
\typeout{** loaded for the language `#1'. Using the pattern for}%
\typeout{** the default language instead.}%
\else
\language=\csname l@#1\endcsname
\fi
#2}}
\providecommand{\BIBdecl}{\relax}
\BIBdecl

\bibitem{devlin2018bert}
J.~Devlin, ``Bert: Pre-training of deep bidirectional transformers for language understanding,'' \emph{arXiv preprint arXiv:1810.04805}, 2018.

\bibitem{baevski2020wav2vec}
A.~Baevski \emph{et~al.}, ``wav2vec 2.0: A framework for self-supervised learning of speech representations,'' \emph{NeurIPS}, 2020.

\bibitem{sadhu2021wav2vec_c}
S.~Sadhu \emph{et~al.}, ``Wav2vec-c: A self-supervised model for speech representation learning,'' \emph{arXiv preprint arXiv:2103.08393}, 2021.

\bibitem{baevski2019vq_wav2vec}
A.~Baevski \emph{et~al.}, ``vq-wav2vec: Self-supervised learning of discrete speech representations,'' \emph{arXiv preprint arXiv:1910.05453}, 2019.

\bibitem{jiang2020speech_simclr}
D.~Jiang \emph{et~al.}, ``Speech simclr: Combining contrastive and reconstruction objective for self-supervised speech representation learning,'' \emph{arXiv preprint arXiv:2010.13991}, 2020.

\bibitem{hsu2021hubert}
W.-N. Hsu \emph{et~al.}, ``Hubert: Self-supervised speech representation learning by masked prediction of hidden units,'' \emph{IEEE/ACM transactions on audio, speech, and language processing}, 2021.

\bibitem{chiu2022best_rq}
C.-C. Chiu \emph{et~al.}, ``Self-supervised learning with random-projection quantizer for speech recognition,'' in \emph{ICML}, 2022.

\bibitem{chen2022wavlm}
S.~Chen \emph{et~al.}, ``Wavlm: Large-scale self-supervised pre-training for full stack speech processing,'' \emph{IEEE Journal of Selected Topics in Signal Processing}, 2022.

\bibitem{chen2024xeus}
W.~Chen \emph{et~al.}, ``Towards robust speech representation learning for thousands of languages,'' \emph{arXiv preprint arXiv:2407.00837}, 2024.

\bibitem{he2022masked}
K.~He \emph{et~al.}, ``Masked autoencoders are scalable vision learners,'' in \emph{CVPR}, 2022.

\bibitem{baevski2022data2vec}
A.~Baevski \emph{et~al.}, ``Data2vec: A general framework for self-supervised learning in speech, vision and language,'' in \emph{ICML}, 2022.

\bibitem{van2017vq_vae}
A.~Van Den~Oord \emph{et~al.}, ``Neural discrete representation learning,'' \emph{Advances in neural information processing systems}, vol.~30, 2017.

\bibitem{babu2021xls}
A.~Babu \emph{et~al.}, ``Xls-r: Self-supervised cross-lingual speech representation learning at scale,'' \emph{arXiv preprint arXiv:2111.09296}, 2021.

\bibitem{chung2021w2v_bert}
Y.-A. Chung \emph{et~al.}, ``W2v-bert: Combining contrastive learning and masked language modeling for self-supervised speech pre-training,'' in \emph{ASRU}.\hskip 1em plus 0.5em minus 0.4em\relax IEEE, 2021, pp. 244--250.

\bibitem{gulati2020conformer}
A.~Gulati \emph{et~al.}, ``Conformer: Convolution-augmented transformer for speech recognition,'' \emph{arXiv preprint arXiv:2005.08100}, 2020.

\bibitem{vaswani2017attention}
A.~Vaswani, ``Attention is all you need,'' \emph{arXiv preprint arXiv:1706.03762}, 2017.

\bibitem{rekesh2023fast}
D.~Rekesh \emph{et~al.}, ``Fast conformer with linearly scalable attention for efficient speech recognition,'' in \emph{ASRU}, 2023.

\bibitem{yang2021superb}
S.-w. Yang \emph{et~al.}, ``Superb: Speech processing universal performance benchmark,'' \emph{arXiv preprint arXiv:2105.01051}, 2021.

\bibitem{barrault2023seamless}
L.~Barrault \emph{et~al.}, ``Seamless: Multilingual expressive and streaming speech translation,'' \emph{arXiv preprint arXiv:2312.05187}, 2023.

\bibitem{radford2023whisper}
A.~Radford \emph{et~al.}, ``Robust speech recognition via large-scale weak supervision,'' in \emph{International conference on machine learning}.\hskip 1em plus 0.5em minus 0.4em\relax PMLR, 2023, pp. 28\,492--28\,518.

\bibitem{puvvada2024less}
K.~C. Puvvada \emph{et~al.}, ``Less is more: Accurate speech recognition \& translation without web-scale data,'' \emph{Interspeech}, 2024.

\bibitem{kahn2020librilight}
J.~Kahn \emph{et~al.}, ``Libri-light: A benchmark for asr with limited or no supervision,'' in \emph{ICASSP}, 2020.

\bibitem{wang2021voxpopuli}
C.~Wang \emph{et~al.}, ``Voxpopuli: A large-scale multilingual speech corpus for representation learning, semi-supervised learning and interpretation,'' \emph{arXiv preprint arXiv:2101.00390}, 2021.

\bibitem{cieri2004fisher}
C.~Cieri \emph{et~al.}, ``The fisher corpus: A resource for the next generations of speech-to-text.'' in \emph{LREC}, vol.~4, 2004, pp. 69--71.

\bibitem{switchboard}
E.~H. John J.~Godfrey, ``Switchboard-1 release 2,'' \url{https://catalog.ldc.upenn.edu/LDC97S62}.

\bibitem{wsj}
J.~S. Garofolo \emph{et~al.}, ``Csr-i (wsj0) complete,'' \url{https://catalog.ldc.upenn.edu/LDC93S6A}.

\bibitem{koh2019nsc}
J.~X. Koh \emph{et~al.}, ``Building the singapore english national speech corpus,'' \emph{Malay}, vol.~20, no. 25.0, pp. 19--3, 2019.

\bibitem{galvez2021people}
D.~Galvez \emph{et~al.}, ``The people's speech: A large-scale diverse english speech recognition dataset for commercial usage,'' \emph{arXiv preprint arXiv:2111.09344}, 2021.

\bibitem{snyder2015musan}
D.~Snyder \emph{et~al.}, ``Musan: A music, speech, and noise corpus,'' \emph{arXiv preprint arXiv:1510.08484}, 2015.

\bibitem{fonseca2017freesound}
E.~Fonseca \emph{et~al.}, ``Freesound datasets: a platform for the creation of open audio datasets.''\hskip 1em plus 0.5em minus 0.4em\relax ISMIR, 2017.

\bibitem{graves2006connectionist}
A.~Graves \emph{et~al.}, ``Connectionist temporal classification: labelling unsegmented sequence data with recurrent neural networks,'' in \emph{ICML}, 2006.

\bibitem{fujita2019end}
Y.~Fujita \emph{et~al.}, ``End-to-end neural speaker diarization with self-attention,'' in \emph{ASRU}, 2019.

\bibitem{desplanques2020ecapa}
B.~Desplanques \emph{et~al.}, ``Ecapa-tdnn: Emphasized channel attention, propagation and aggregation in tdnn based speaker verification,'' \emph{arXiv preprint arXiv:2005.07143}, 2020.

\bibitem{horiguchi2022eend_eda}
S.~H. et~al., ``Encoder-decoder based attractors for end-to-end neural diarization,'' \emph{IEEE/ACM Transactions on Audio, Speech, and Language Processing}, vol.~30, pp. 1493--1507, 2022.

\bibitem{horiguchi2022eend_gla}
S.~Horiguchi \emph{et~al.}, ``Online neural diarization of unlimited numbers of speakers using global and local attractors,'' \emph{IEEE/ACM Transactions on Audio, Speech, and Language Processing}, vol.~31, pp. 706--720, 2022.

\bibitem{park2022msdd}
T.~J. Park \emph{et~al.}, ``Multi-scale speaker diarization with dynamic scale weighting,'' \emph{arXiv preprint arXiv:2203.15974}, 2022.

\bibitem{park2024sortformer}
\BIBentryALTinterwordspacing
T.~Park \emph{et~al.}, ``Sortformer: Seamless integration of speaker diarization and asr by bridging timestamps and tokens,'' \emph{arXiv preprint arXiv:2409.06656}, 2024. [Online]. Available: \url{https://arxiv.org/abs/2409.06656}
\BIBentrySTDinterwordspacing

\bibitem{noroozi2024stateful}
V.~Noroozi \emph{et~al.}, ``Stateful conformer with cache-based inference for streaming automatic speech recognition,'' in \emph{ICASSP}, 2024.

\bibitem{commonvoice:2020}
R.~Ardila \emph{et~al.}, ``Common voice: A massively-multilingual speech corpus,'' in \emph{LREC}, 2020, pp. 4211--4215.

\bibitem{pratap2020mls}
V.~Pratap \emph{et~al.}, ``Mls: A large-scale multilingual dataset for speech research,'' \emph{arXiv preprint arXiv:2012.03411}, 2020.

\bibitem{o2021spgispeech}
P.~K. O'Neill \emph{et~al.}, ``Spgispeech: 5,000 hours of transcribed financial audio for fully formatted end-to-end speech recognition,'' \emph{arXiv preprint arXiv:2104.02014}, 2021.

\bibitem{iranzo2020europarl}
J.~Iranzo-S{\'a}nchez \emph{et~al.}, ``Europarl-st: A multilingual corpus for speech translation of parliamentary debates,'' in \emph{ICASSP}, 2020.

\bibitem{panayotov2015librispeech}
V.~Panayotov \emph{et~al.}, ``Librispeech: an asr corpus based on public domain audio books,'' in \emph{ICASSP}, 2015.

\bibitem{stt_en_fastconformer_transducer_xlarge}
{NVIDIA}, ``Nemo english fastconformer-rnnt asr model,'' \url{https://catalog.ngc.nvidia.com/orgs/nvidia/teams/nemo/models/megatronnmt_any_en_500m}.

\bibitem{MegatronNMTAnyEn2024}
NVIDIA, ``Megatron multilingual translation model,'' \url{https://catalog.ngc.nvidia.com/orgs/nvidia/teams/nemo/models/megatronnmt_any_en_500m}.

\bibitem{mexpresso}
META, ``mexpresso (multilingual expresso),'' \url{https://huggingface.co/facebook/seamless-expressive#mexpresso-multilingual-expresso}.

\bibitem{conneau2023fleurs}
A.~Conneau \emph{et~al.}, ``Fleurs: Few-shot learning evaluation of universal representations of speech,'' in \emph{SLT}, 2023.

\bibitem{ami}
U.~of~Edinburgh, ``The ami corpus,'' \url{https://www.openslr.org/16/}.

\bibitem{icsi}
------, ``The icsi meeting corpus,'' \url{https://groups.inf.ed.ac.uk/ami/icsi/}.

\bibitem{ryant2020third}
N.~Ryant \emph{et~al.}, ``Third dihard challenge evaluation plan,'' \emph{arXiv preprint arXiv:2006.05815}, 2020.

\bibitem{voxconv}
J.~S. Chung \emph{et~al.}, ``Spot the conversation: speaker diarisation in the wild,'' \emph{arXiv preprint arXiv:2007.01216}, 2020.

\bibitem{fu2021aishell}
Y.~Fu \emph{et~al.}, ``Aishell-4: An open source dataset for speech enhancement, separation, recognition and speaker diarization in conference scenario,'' \emph{arXiv preprint arXiv:2104.03603}, 2021.

\bibitem{callhome2}
Kaldi, ``Kaldi x-vector recipe v2,'' \url{https://github.com/kaldi-asr/kaldi/tree/master/egs/callhome_diarization/v2}.

\bibitem{callhome}
M.~Przybocki \emph{et~al.}, ``2000 nist speaker recognition evaluation,'' \url{https://catalog.ldc.upenn.edu/LDC2001S97}.

\bibitem{doddington2000nist}
G.~R. Doddington \emph{et~al.}, ``The nist speaker recognition evaluation--overview, methodology, systems, results, perspective,'' \emph{Speech communication}, vol.~31, no. 2-3, pp. 225--254, 2000.

\bibitem{nist}
{NIST}, ``Nist speaker recognition evaluation (sre),'' \url{https://www.nist.gov/itl/iad/mig/speaker-recognition}.

\bibitem{park2023property}
T.~J. Park \emph{et~al.}, ``Property-aware multi-speaker data simulation: A probabilistic modelling technique for synthetic data generation,'' \emph{arXiv preprint arXiv:2310.12371}, 2023.

\bibitem{bastianelli2020slurp}
E.~Bastianelli \emph{et~al.}, ``Slurp: A spoken language understanding resource package,'' \emph{arXiv preprint arXiv:2011.13205}, 2020.

\bibitem{wang2021fine}
Y.~Wang \emph{et~al.}, ``A fine-tuned wav2vec 2.0/hubert benchmark for speech emotion recognition, speaker verification and spoken language understanding,'' \emph{arXiv preprint arXiv:2111.02735}, 2021.

\bibitem{arora2022espnet}
S.~Arora \emph{et~al.}, ``Espnet-slu: Advancing spoken language understanding through espnet,'' in \emph{ICASSP}, 2022.

\bibitem{whetten2024open}
R.~Whetten \emph{et~al.}, ``Open implementation and study of best-rq for speech processing,'' \emph{arXiv preprint arXiv:2405.04296}, 2024.

\bibitem{seo2022integration}
S.~Seo \emph{et~al.}, ``Integration of pre-trained networks with continuous token interface for end-to-end spoken language understanding,'' in \emph{ICASSP}, 2022.

\bibitem{huang2023leveraging}
H.~Huang \emph{et~al.}, ``Leveraging pretrained asr encoders for effective and efficient end-to-end speech intent classification and slot filling,'' \emph{Interspeech}, 2023.

\end{thebibliography}

\end{document}